\documentclass[prd,aps,floatfix,superscriptaddress,nofootinbib,twocolumn,10pt,English]{revtex4}

\usepackage{amssymb,amsmath}
\usepackage{graphicx}
\usepackage[percent]{overpic}
\usepackage{overpic}
\usepackage{color}
\usepackage[colorlinks=true]{hyperref}
\usepackage{braket}

\usepackage{soul}
\setstcolor{red}

\def\ba{\begin{eqnarray}}
\def\ea{\end{eqnarray}}

\begin{document}  


\title{Gravitational floating orbits around hairy black holes} 
{
\author{Jun Zhang}
\email{jun.zhang@imperial.ac.uk}
\affiliation{Theoretical Physics, Blackett Laboratory, Imperial College, London, SW7 2AZ, U.K. }
\affiliation{Department of Physics and Astronomy, York University, Toronto, Ontario, M3J 1P3, Canada}
\affiliation{Perimeter Institute for Theoretical Physics, Waterloo, Ontario N2L 2Y5, Canada }
\author{Huan Yang}
\email{hyang@perimeterinstitute.ca}
\affiliation{Perimeter Institute for Theoretical Physics, Waterloo, Ontario N2L 2Y5, Canada }
\affiliation{University of Guelph, Guelph, Ontario N2L 3G1, Canada}

\begin{abstract} 
We show that gravitational floating orbits may exist for black holes with rotating hairs. These black hole hairs could originate from the superradiant growth of a light axion field around the rotating black holes. If a test particle rotates around the black hole, its tidal field may resonantly trigger the dynamical transition between a co-rotating state and a dissipative state of the axion cloud. A tidal bulge is generated by the beating of modes, which feeds angular momentum back to the test particle. Following this mechanism, an extreme-mass-ratio-inspiral (EMRI) system, as a source for LISA, may face delayed merger as the EMRI orbit stalls by the tidal response of the cloud, until the cloud being almost fully dissipated. If the cloud depletes slower than the average time separation between EMRI mergers, it may lead to interesting interaction between multiple EMRI objects at comparable radii. Inclined EMRIs are also expected to migrate towards the black hole equatorial plane due to the tidal coupling and gravitational-wave dissipation.
Floating stellar-mass back holes or stars around the nearby intermediate-mass black holes may generate strong gravitational-wave emission detectable by LISA.
\end{abstract}

\maketitle 

\section{Introduction}

Black Hole (BH) No Hair Theorem states that any stationary black hole in Einstein-Maxwell theory can be characterized by its mass, spin and electric charge, which is possible to be tested with BH spectroscopy in the Advanced LIGO (Laser Interferometric Gravitational-Wave Observatory) era  \cite{berti2016spectroscopy,yang2017black,brito2018black,Thrane_2017}.
If additional  { bosonic} fields are allowed in the setup, they may grow exponentially according to the BH superradiance \cite{Detweiler:1980uk,zouros1979instabilities} and saturate onto  quasi-stationary configurations \cite{east2017superradiant,east2017superradiant2}.
In particular, these hair fields (such as the QCD Axion \cite{weinberg1978new}, dark photons \cite{holdom1986two,cicoli2011testing} and string axiverse \cite{arvanitaki2010string}) around BHs may serve as Dark Matter candidates, and depending on their mass range, they could be dynamically important to the spin evolution of isolated BHs. The rotation of these fields may also generate continuous gravitational waves (GWs) that lie in the detection band of LIGO or LISA (Laser Interferometric Space Antenna) \cite{arvanitaki2015discovering,baryakhtar2017black,Brito:2017zvb,brito2017stochastic}.

The rotating cloud can carry a significant fraction of energy/angular momentum (AM) of the host BH. Since the BH area generally increases following the superradiant growth of the cloud \cite{east2017superradiant2}, while interacting with an external agent, the cloud AM would not be entirely re-absorbed by the host BH (e.g., through the tidally-induced cloud depletion discussed in \cite{Baumann:2018vus}), or its horizon area would decrease. As a result, the external agent must acquire part of the cloud energy/AM during the interaction process. This AM transfer may give rise to {\it gravitational floating orbits} of a test particle, in which case the GW damping of the orbital energy and AM is balanced by the gravitational interaction with the cloud. Such orbits are first conjectured in \cite{press1972floating},  based on the observation that the horizon AM flux generated by a test particle orbiting around a rotating BH could be negative due to the superradiance effect. However, for Kerr BHs the AM gain from horizon is universally weaker than the loss due the GW radiation at infinity, which means that there is no gravitational floating orbit in Kerr spacetime. On the other hand, if the particle also couples to a massive scalar field  besides the gravitational interaction, it has been shown \cite{Cardoso:2011xi,Ferreira:2017pth} \footnote{The  argument of \cite{Ferreira:2017pth} is drawn in analogy to planetary systems, and a complete analysis including the backreaction on the cloud is required to prove the existence of  positive AM transfer during resonances. The resonance studied here operates at lower frequency, and is still valid for complex scalar field.} that the scalar wave radiation can balance the GW radiation, and lead to floating orbits given suitable scalar field mass and coupling strength.

\begin{figure}[tbp]
\centering 
\includegraphics[width=0.4\textwidth]{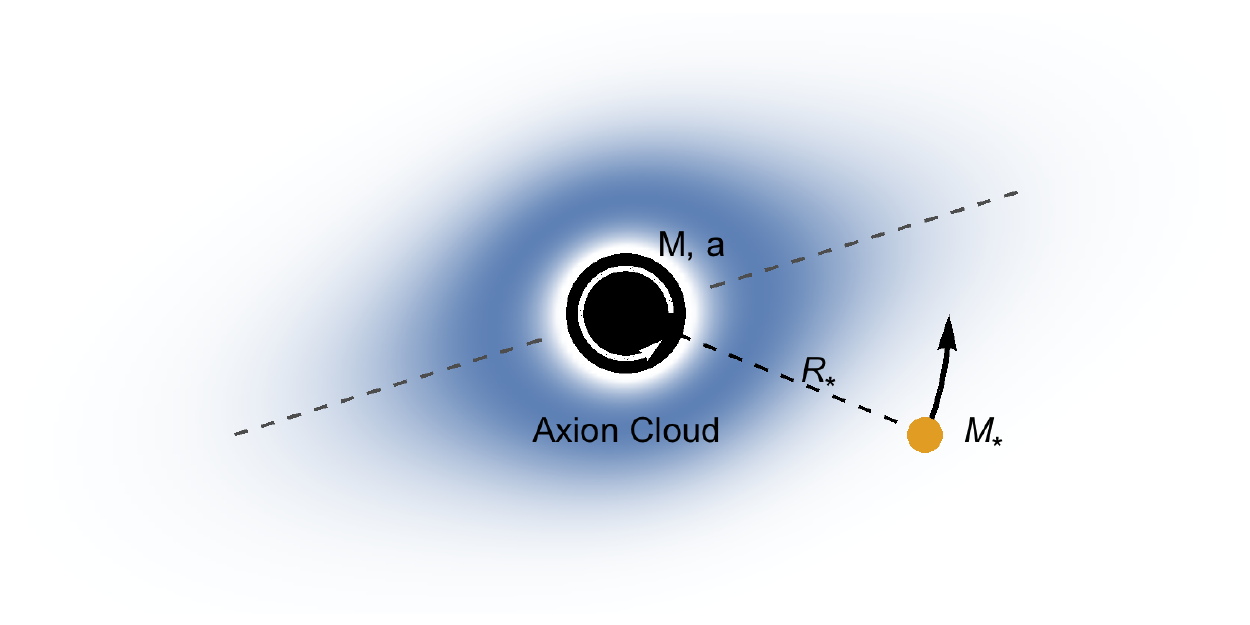} 
\caption{A host BH of mass $M$ and dimensionless spin $a$ dressed with an axion cloud and companied by a star of mass $M_*$ at $R_*$. The axion cloud develops by superradiance, and is quasi-stationary without the companion star. Through the tidal interaction  with the star, a bulge of the cloud develops, which {\it leads} the motion of the star. In return, the star extracts AM from the BH and the cloud to compensate the AM loss due to GW radiation. The star would float at this orbit until the whole cloud is depleted.} \label{fig:cartoon}
\end{figure}

In this paper we show that indeed the tidal interaction between a rotating cloud and a test particle could support gravitational floating orbits, without assuming additional axion field-matter interactions. Physically the test particle tidally deforms the cloud. Due to the cloud dissipation, there is a phase difference between the particle's orbit and the tidal bulge. Unlike the tidal interactions commonly seen in binary stars, the tidal bulge in the cloud actually leads the test particle's motion, and consequently AM transfers from the cloud to the particle. 

We examine this cloud energy/AM transfer mechanism in the context of EMRIs, which are important sources for LISA. We find that for a range of EMRI mass ratio and axion mass, the EMRI orbit stalls at finite radius until the axion cloud is depleted. Notice that this process could take longer than the inspiralling time of the EMRI, which implies interesting astrophysical  effects. Unless  { specified}, we set $G = c =\hbar =1$.

\section{Toy model}   

We first illustrate the physical mechanism using a two-mode model. Let us consider a BH with mass $M$ and dimensionless spin $a$, dressed with axion cloud. Like the electron cloud in a hydrogen atom, the axion cloud also possesses a tower of eigenmodes,  denoted by $\ket{nlm}$ with $\{n, l, m\}$ being the principal, orbital, and magnetic ``quantum number" respectively. In particular, a mode with $m>0$ is growing if its eigenfrequency $\omega_{nlm} < m \Omega_H$, with $\Omega_{H}$ being the horizon frequency of the BH, and a mode with $m \leqslant 0$ is always decaying. The toy model involves a growing mode and a decaying mode, e.g. $\ket{211}$ and $ \ket{21-1}$. At linear level, these two modes evolve independently, but could become coupled in the presence of an external tidal field provided by a companion star. As depicted in Fig.~\ref{fig:cartoon}, the star has a mass $M_*$, and for simplicity we assume it is co-orbiting with the BH in a quasi-circular orbit of radius $R_*$ on the equator. 

In the interaction picture, the wavefunction of the axion cloud is a linear combination of  two modes
\ba
\ket{\psi(t)} = c_g(t) \ket{\psi_g} + c_d(t) \ket{\psi_d} ,
\ea
where $c_g$ and $c_d$ are the time-dependent amplitudes, with subscripts $g$ and $d$ denoting the growing and decaying mode respectively. Initially, $\ket{\psi}$ is normalized as $\bra{\psi(0)}\mu\ket{\psi(0)} = M_c$, where $\mu$ is the mass of the scalar field and $M_c$ is the mass of the cloud. The mass of the saturated cloud depends on the initial spin of the BH, and should be determined by numerical simulations. The theoretical upper limit of super radiance extraction is given by $M_c/M < 0.29$ \cite{Christodoulou:1970wf}. However, recent simulations show that the cloud can store at most $\sim 10\%$ of the BH's mass \cite{east2017superradiant,Herdeiro:2017phl}. In this paper, we assume the initial BH spin is close to maximal in which case the mass of cloud can be estimated as $M_c \sim \alpha M$ for $\alpha \ll 1$ (See Eq.(27) in reference \cite{Brito:2017zvb}), where we have defined
\ba\label{eq:alpha}
\alpha \equiv \mu M \simeq 0.1 \left(\frac{M}{10 M_\odot}\right)\left(\frac{\mu}{10^{-12}\text{eV}}\right).
\ea
In the non-relativistic limit, the coefficients $\mathbf c \equiv \left(c_g,\, c_d\right)^{T}$ satisfy the Schr\"odinger equation $i\, {\rm d} \mathbf c/{\rm d}t = \mathbf H_I \mathbf c$ with
\begin{align}
 \mathbf H_I =    \begin{pmatrix} 0  &  \eta \, e^{- i  \left( \Delta m \Omega - \Delta\omega \right) t }  &  \\  \eta \, e^{+  i  \left( \Delta m \Omega - \Delta\omega \right) t }  &  -i\Gamma \end{pmatrix}, \label{equ:Schrodinger}
\end{align}
where  $\Delta m \equiv m_g-m_d$, $\Omega$ is the orbital frequency of the companion star, 
$\Delta\omega \equiv  {\rm Re} \left[\omega_{g}-\omega_{d}\right]$
is the energy split of the two modes, and $\Gamma = -{\rm Im}\left[\omega_{d}\right]$ is the damping rate of the $\ket{21-1}$ mode. Following \cite{Detweiler:1980uk}, we take $\Gamma \simeq \mu\alpha^{9}/6$ for $\alpha \ll 1$ in the slowly rotating limit as the rotation of the BH has been slowed down by the supperradiance.
In the Newtonian limit,  the energy split is given by $\Delta\omega \simeq a \alpha^5 \mu/6$ and the mode coupling (off-diagonal terms) is induced by the quadrupole tidal perturbations, with $\eta = 9\alpha^{-3}q M^2 /R_*^3$ and $q \equiv M_*/M$ \cite{Baumann:2018vus}. We assume that initially the cloud is saturated, purely consisting of the $\ket{211}$ mode, i.e. $c_g(0) = 1$ and $c_d(0)=0$. By dynamically evolving $c_g(t)$ and $c_d(t)$, we  find that the wavefunction oscillates between the modes with Rabi frequency
\ba
\omega_{\rm R} = \sqrt{\eta^2 + \left(\Delta\omega - \Delta m \Omega\right)^2/4}
\ea
due to tidal coupling, and a resonance occurs when the orbital frequency matches the energy split $\Omega \sim \Delta\omega/\Delta m$.

Close to the resonance, the cloud  loses AM to the BH due to the excitation of the decaying mode. In fact, the AM flux at the horizon can be estimated as \cite{Dolan:2007mj}
\ba\label{eq:transferrate}
\left< \left.\frac{dL}{dt}\right|_{H}\right> = \frac{-2\Gamma}{T}\int_0^{T}dt \, m_d c_d^*c_d = -m_d \Gamma \frac{\eta^2}{2\omega_{\rm R}^2},
\ea
where the time average is taken for many Rabi oscillation periods. Notice that as the decaying mode is losing ``negative" AM to the BH, the BH AM changes as $\left< dL_{BH}/dt\right> = - \left< \left.dL/dt\right|_{H}\right>$. On the other hand, the AM of the cloud can be calculated as:
\ba
L_c = m_g c_g^* c_g + m_d c_d^* c_d,
\ea
which implies an averaging AM change rate $\left< dL_c/dt \right> =  -m_g\Gamma \eta^2/2\omega_{\rm R}^2$ up to the linear order in $\Gamma$, as $\Gamma$ is much smaller than $\Delta \omega$. As the total AM is conserved, the AM of the companion star changes as
\ba\label{eq:transferrate}
\left< \frac{dL_*}{dt}\right> = -\left< \frac{dL_{BH}}{dt}\right>-\left< \frac{dL_c}{dt}\right> =\Delta m \Gamma\frac{\eta^2}{2\omega_{\rm R}^2}.
\ea
The AM gained by the companion star also can be computed by considering the back reaction from the deformed cloud to the star. The tidal density deformation is
\ba
\left< \delta \rho \right> \simeq \frac{\eta}{2\omega_{\rm R}^2}\sqrt{4\omega_{\rm R}^2+\Gamma^2} \cos\left[\Delta m \left(\phi - \Omega t\right) + \delta\phi \right] \rho_{\times},
\ea
where $\rho_{\times} =\exp\left[i \Delta m \phi\right] \psi^*_g \psi_d$, and $\sin\delta\phi \equiv \Gamma/\sqrt{4\omega_{\rm R}^2+\Gamma^2}$. The deformed density induces 
an additional tangential gravitational acceleration for the star
consistent with Eq.~\eqref{eq:transferrate}.

It is possible that AM loss of the star due to its GW radiation is compensated by the gain from from the cloud. As a result, the orbital decay stalls because of the AM transfer. Using the balance condition, we find that the companion star  floats at an orbit frequency $\Omega_F \equiv  (1-\epsilon)\Delta\omega/2$ with 
\ba\label{eq:Deltax}
\epsilon \simeq \frac{3^{5/3}\sqrt{5}}{8} \alpha^{5/6} \sqrt{M_c/\alpha M}\, ,
\ea 
until the cloud depletes completely.

Although the above discussion is for a circular orbit, we expect  similar results hold for an elliptical orbit. Comparing to a circular orbit, the tidal perturbation of an elliptical orbit can be decomposed to harmonics with multiples of orbital frequencies. As a result, the coupling between two certain modes can be written as a Fourier series that is periodic with the orbital period. However, the coupling between the two modes (and hence the angular momentum transfer) is only dominated by the harmonic whose frequency matches the resonance frequency, hence introducing the  frequency locking. Physically the induced tidal bulge of the cloud still transfers angular momentum along its spin axis to ensure that the floating object keep the same orbital frequency, although the orbital eccentricity may decrease overtime due to the gravitational-wave radiation.

 \vspace{0.2cm}
 
\section{BH Perturbation} 
We now perform a BH perturbation calculation on hairy BH systems to obtain details in the fully relativistic regime. 
If the axion's Compton wavelength is much larger than the size of the BH, although trapped, the support of the axion density profile is away from the BH, which justifies an approximate Newtonian treatment. For more general axion parameters, a BH perturbation analysis is necessary.
\begin{figure}[tbp]
\centering 
\includegraphics[height=0.27\textwidth]{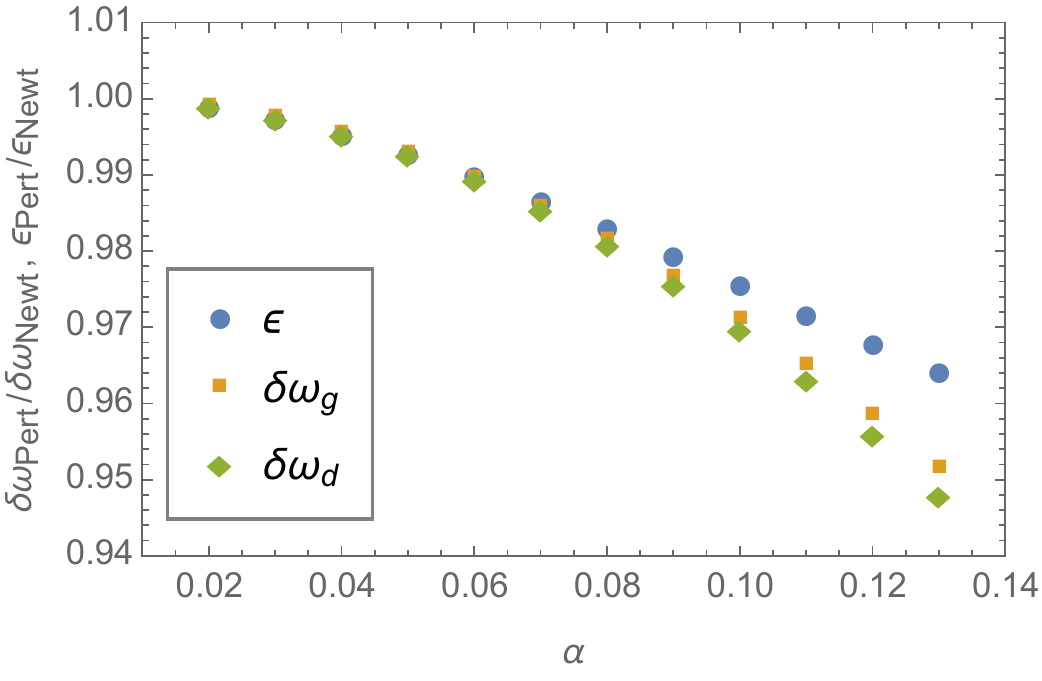} 
\caption{Comparison between the BH perturbation and the Newtonian treatment. $\delta \omega_g$ and $\delta \omega_d$ are the frequency shifts of the $\ket{211}$ and $\ket{21-1}$ modes generated by the tidal field of companion star with $M_*=10^{-5} M$. $\epsilon$ is defined by $\Omega_F\equiv \left(1-\epsilon\right) \Delta\omega/2$ with $\Omega_F$ being the floating orbit frequency. We assume that the mass of the axion cloud is $\alpha M$. In the Newtonian treatment, $\delta \omega_g = \delta \omega_d =-3\alpha^{-3}M_*M/R_*^3$ and $\epsilon$ can be estimated by Eq.~(\ref{eq:Deltax}).
} \label{fig:comparison}
\end{figure}

The cloud is still assumed to be fully grown to its saturation limit in the absence of a tidal perturber. We approximate the density distribution of a fully grown cloud according to the eigenmode wave function  \footnote{The exact solution can be found in \cite{herdeiro2015construction}, and the numerical solution is presented in \cite{east2017superradiant2}. Eigenmode is a good approximation as the cloud energy is generally small comparing to the BH mass.}. The evolution of a scalar field $\Psi$ with mass parameter $\mu$ on a perturbed Kerr background $g=g_{\rm Kerr}+h$ can be described by
\begin{align}\label{eqboxop}
&(\Box_{g} +\mu^2)\Psi =0\nonumber \\
&\approx  (\Box_{\rm Kerr}+\mu^2) \Psi- \frac{1}{\sqrt{-g_{\rm Kerr}}} \partial_\mu \left( h^{\mu \nu} \sqrt{-g_{\rm Kerr}} \partial_\nu \Psi \right) \notag \\ 
&+ \frac12 g_{\rm Kerr}^{\mu \nu} \left(\partial_\mu h^{\rho}{}_\rho \right)  \partial_\nu \Psi\, 
\equiv \left [ \Box_{\rm Kerr}+\mu^2+\frac{1}{\Sigma} \mathcal{H}(h) \right ] \Psi\,,
\end{align} 
where $\Sigma := r^2+a^2\cos^2\theta$, and the operator $\mathcal{H}(.)$ is linear in its argument. We adopt the tidal-deformation metric $h$ from \cite{Poisson:2014gka}, for a slowly rotating black hole with a companion star.

The above wave equation can also be written as
$\left [ \Box_{\rm Kerr}+\mu^2\right ] \Psi={\rm S}$, with $S \equiv -\frac{1}{\Sigma} \mathcal{H}(h) \Psi$. Formally its solution is
\begin{align}
\Psi = \int d^4 x' G(x, x') {\rm S}(x')\,,
\end{align}
with the Green function $G(x,x')$ satisfying
\begin{align}
\left [ \Box_{\rm Kerr}+\mu^2\right ] G(x,x') =  \delta^{(4)}(x-x')\,.
\end{align}

According to the discussion in \cite{yang2014scalar} and taking into account the fact that $\mu \neq 0$, the Green function can be decomposed into two parts in the frequency domain: $G = G_{\rm dir}+G_{\rm QNM}$. The ``direct" part $G_{\rm dir}$ generates the propagating waves that travel to spatial infinity or into the BH horizon, with the explicit form unknown. It usually disappears fast for transient sources. The QNM part $G_{\rm QNM}$ generates QNM ringing that we study here. For Kerr BHs it can be expressed as
\begin{align}
\label{eqqnmgreen}
G_{\rm QNM}&(x,x')= -\frac{2}{ \sqrt{r^2+a^2} \sqrt{r'^2+a^2}}  \notag \\ &
\times {\rm Re}\Biggl[\sum_m e^{i m(\phi-\phi')}\sum_{l, n} Y_{lm}(\theta)Y^*_{lm}(\theta')  \notag \\ &
\qquad \times \mathcal{A}_{nlm}u_{\rm in}(r)u_{\rm in}(r')e^{-i\omega_{nlm} (t-t')}\Biggr].
\end{align}
with $Y_{lm}(\theta)$ being the spheroidal harmonics, $\omega_{nlm}$ being the QNM frequency with spherical index $l,m$ and radial overtone $n$, $\mathcal{A}_{nlm}$ equal to $[2 \omega C^+_{\omega lm} \partial_\omega C^-_{\omega lm}]^{-1}_{\omega=\omega_{nlm}}$ and the scattering coefficient $C^{\pm}_{\omega lm}$ given in \cite{yang2014scalar}. The wave function  $u_{\rm in}$ ($nlm$ sub-indices abbreviated) is just $(r^2+a^2)^{1/2}R_{\rm in}$, where $R_{\rm in}$ satisfies the radial Teukolsky equation (c.f. \cite{yang2014scalar}) and is solved in \cite{Dolan:2007mj} with the bound state boundary condition. Focusing on the QNM Green function, we write the QNM sum as
\begin{align}
\Psi_{\rm QNM} &= \sum_{n l m } A_{nlm}(t) e^{-i \omega_{nlm} t} R_{\rm in}(r) Y_{lm}(\theta) e^{i m \phi} \nonumber \\
& \equiv \sum_{nlm} A_{nlm}(t) e^{-i \omega_{nlm} t} \psi_{nlm} \nonumber \\
& = \int d^4 x' G_{\rm QNM}(x,x') S(x')\,.
\end{align}

As shown in the toy model, mode coupling becomes significant only when the frequency of the perturbation matches the energy split, which allows us to restrict ourselves to a two-mode subspace, as they are the main excitations given a certain external perturbation. The mode equations of motion are \cite{yang2015turbulent,yang2015coupled}
\ba\label{eqm}
&&\frac{\dot{A_g}}{2\mathcal{A}_g}  \approx A_g \langle \psi_g |  \mathcal{H}(h) | \psi_g \rangle + A_d e^{i \Delta \omega\, t}\langle \psi_g |  \mathcal{H}(h) | \psi_d \rangle ~~~~~~\\
&&\frac{\dot{A_d}}{2\mathcal{A}_d}  \approx A_g e^{-i \Delta \omega \, t} \langle \psi_d|   \mathcal{H}(h) | \psi_g \rangle + A_d \langle \psi_d | \mathcal{H}(h) | \psi_d \rangle\,.\nonumber 
\ea
 The inner product is defined as 
\begin{align}\label{eq:inner}
\langle \psi_{nlm} | \eta \rangle \equiv \int dr \int d\phi \int d\theta \sin\theta R_{in}Y_{lm}^* e^{-i m \phi} \, \eta\,,
\end{align}
where the integral over $r$ direction is regularized to remove apparent singularity of the integrand near the horizon, following \cite{Leaver:1986gd, Detweiler:1979xr}. According to the discussion in \cite{zimmerman2015quasinormal,mark2015quasinormal,yang2016plasma}, $2 \mathcal{A}$ can be alternatively evaluated as $-i \langle \psi | \partial_\omega \tilde{\Box} | \psi \rangle^{-1}$, with $\tilde{\Box}$ being $\Box_{\rm Kerr}$ in frequency space.
Notice that the diagonal terms in Eq.~\eqref{eqm} generate constant frequency shifts of eigenmodes. The off-diagonal terms generate the transition between modes, and consequently the AM transfer.
\begin{figure}[tbp]
\centering 
\includegraphics[height=0.23\textwidth]{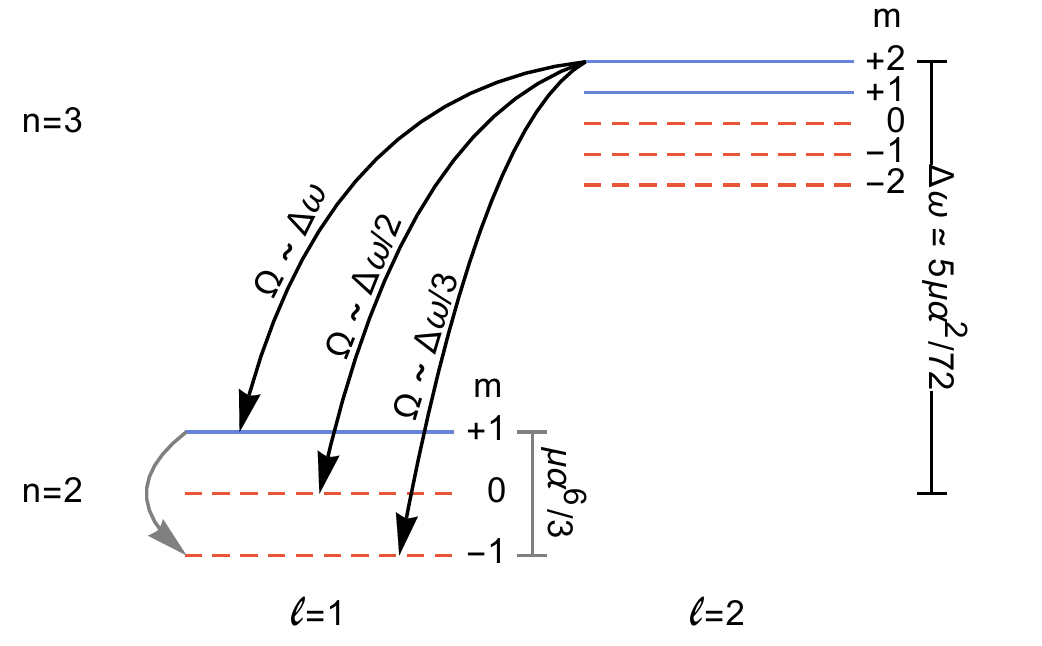} 
\caption{The eigenfrequencies of modes with $n=2, l=1$ and $n=3, l=2$. The blue solid lines are the growing modes and the red dashed lines are the decaying modes. Note that when the $\ket{322}$ mode saturates, the $\ket{211}$ mode also becomes a decaying mode. The arrows show the possible resonances  associated with a co-rotating companion star. } \label{fig:levels}
\end{figure}

Taking $\ket{211}$ and $\ket{21-1}$ as an example, we calculate the frequency shift generated  by a companion star and the AM flux at the horizon.  The  comparison to that from the Newtonian treatment are shown in Fig.~\ref{fig:comparison}. We find that the results start to deviate from their Newtonian counterpart when $\alpha > 0.1$.

\section{Floating Orbits}

Given the superradiance efficiency of each mode, the axion cloud around an astrophysical BH is possibly dominated by a saturated mode with $n-1=m=l$, where $l =1, 2, 3...$ depends on the formation time of the BH \cite{Arvanitaki:2010sy}. In principle, a growing mode could couple to many decaying modes simultaneously. However, a resonance happens when the orbital frequency $\Omega$ is approximately $(\omega_g - \omega_d)/(m_g-m_d)$. This condition has two implications. First, the saturation condition requires that  $m_g - m_d > 0$, thus a co-rotating companion star ($\Omega > 0$) can only couple a growing mode  to a lower-frequency decaying modes, and vice versa. According to Eq. (\ref{eq:transferrate}), a companion star only gain positive AM, therefore a floating orbit does not exist for counter-rotating stars ($\Omega < 0 $). Secondly, at any orbital frequency, a parent growing mode only efficiently couples to one daughter decaying mode, because the width of the resonance band, characterized by $\epsilon \Delta\omega$ with $\epsilon \sim \alpha^{7/2}$ for the $\ket{322}$ mode for example, is much smaller comparing to the frequency separation between modes which is of the order of $\Delta\omega$.  According to Eq.~(\ref{eq:transferrate}), the AM transfer rate depends on the decay rate of the decaying mode, which is proportional to $\alpha^{4l_d+5}$. Therefore, an efficient transfer is usually provided by the mode with lower $l_d$. Given the inner product defined in Eq. (\ref{eq:inner}), we  find that growing modes always couple to the $\ket{n1-1}$ modes though a tidal perturbation with $l_* = l_g+1$. 
\begin{figure}[tbp]
\centering 
\includegraphics[height=0.23\textwidth]{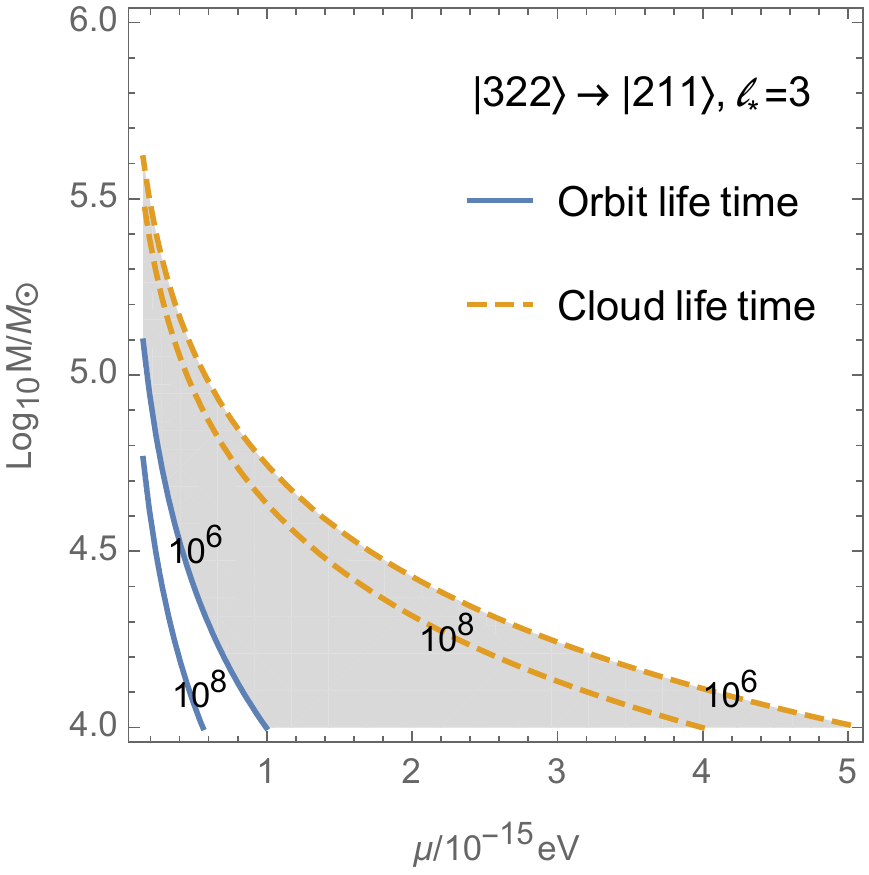} \,
\includegraphics[height=0.23\textwidth]{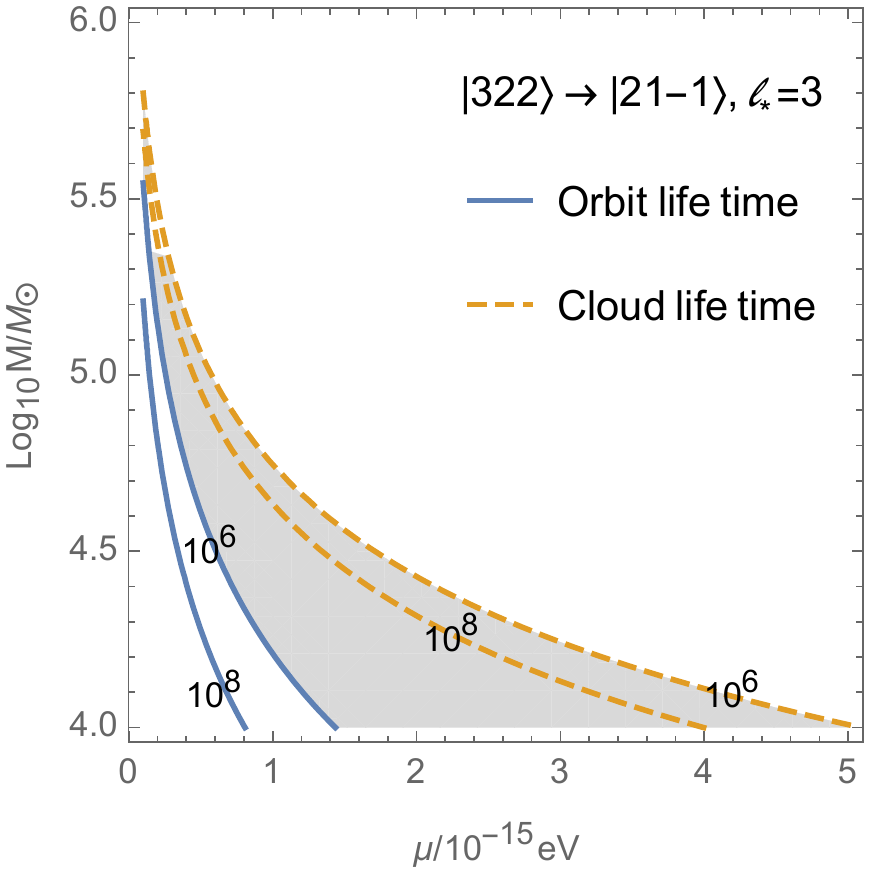} 
\caption{The parameter space of having a floating orbit caused by the $\ket{322}$ mode coupled to the $\ket{211}$ mode (left panel) and the $\ket{21-1}$ (right panel). The orbit is assumed to be on the equator. The blue solid contours show the GW damping time of the orbits if there is no axion cloud, and the orange dashed contours show the lifetime of the axion cloud. The shaded region shows the parameters to having a floating orbit with a lifetime greater than $10^6$ years. Numbers are displayed in  the unit of years.} \label{fig:para}
\end{figure}

The eigenfrequencies of the first two growing modes and the relevant daughter modes are shown in Fig~\ref{fig:levels}. For the $\ket{211}$ mode as a dominant mode, a floating orbit can exist only by its coupling to the $\ket{21-1}$ mode (as shown by the grey arrow in the Fig.~\ref{fig:levels}) 
. However, the frequency difference between these two modes is $M\Delta \omega \sim \alpha^7/3$, which is so small that the associated floating orbit has a radius of $R_F \sim 3^{2/3} M/\alpha^{14/3}$, far away from the central BH. As a result, the life time of such orbit $\sim 5 \left(8 M \Omega \right)^{-8/3} M^2 \simeq 6\times 10^{-4} \alpha^{-56/3} (M/10^5M_\odot)^2\, {\rm yr}$, even without floating, is longer than the lifetime of the cloud $\sim 10^{-6} \alpha^{-15} (M/10^5M_\odot)\, {\rm yr}$. The existence of the axion cloud does not alter the orbit decay significantly, and is of minimal astrophysical relevance.

For the $\ket{322}$ mode, a floating orbit exists by the coupling to the $\ket{211}$ mode or to the $\ket{21-1}$ mode via an octopole tidal perturbation ($l_*=3$). For coupling between modes with different $n$, the floating orbital frequency scales as $M\Omega \sim \alpha^3$. Therefore the orbital radius is comparable to the radius of the axion cloud $R_c \sim M\alpha^{-2}$, namely the companion star is within the axion cloud. Nevertheless, the perturbation method still applies since the mass of the companion star is much smaller than that of the axion cloud. Using the balance condition, we find that, for the coupling to the $\ket{21-1}$ mode, the orbit floats at $\Omega_F = (1-\epsilon)  \Delta\omega /3$ with $\epsilon \simeq 2.6 \alpha ^{7/2}\sqrt{M_c/\alpha M}$, which is far away from other resonance frequencies, such as, $\Delta\omega$ or $\Delta\omega/2$ for the $\ket{211}$ or $\ket{210}$ mode respectively. In Fig.~\ref{fig:para}, we present viable physical parameters that allow floating orbits associated with the $\ket{322}$ mode, with the orbit assumed to lie on the equator of the BH. The requirements are two-fold. 
Based on the EMRI rate in 
\cite{gair2017prospects},
 the orbit lifetime $\tau$ (GW damping timescale, Blue Solid lines) of the nearest perturber should be $\mathcal{O}(10^6)$ yrs. On the other hand, at the time of interest, the cloud's dominant mode depends on the BH's formation history and age, as each unstable mode only survives for a finite time due to GW radiation \cite{Yoshino:2013ofa,Brito:2017zvb}. For the $\ket{322}$ mode, the BH's age should not exceed the mode lifetime (Orange Dashed lines). 

At the end of this section, we would like to briefly comment on the stability of floating orbit. As we discussed, the back reaction of GW radiation exerts a negative torque on the star, while the cloud exerts a positive torque that peaked at the resonance frequency. As shown in Fig.~\ref{fig:torque}, there will be two orbits where these two torques balance. Physically as the EMRI orbit decays, the orbit hits the outer balance point outside the resonance peak first. In this case, an inward radial perturbation on the orbit will lead to a larger torque that push the star outward, and vice verse. Therefore, the outer floating orbit is stable. For the same reason, the inner floating orbit is unstable.
\begin{figure}[tbp]	
\centering
\includegraphics[width=0.365\textwidth]{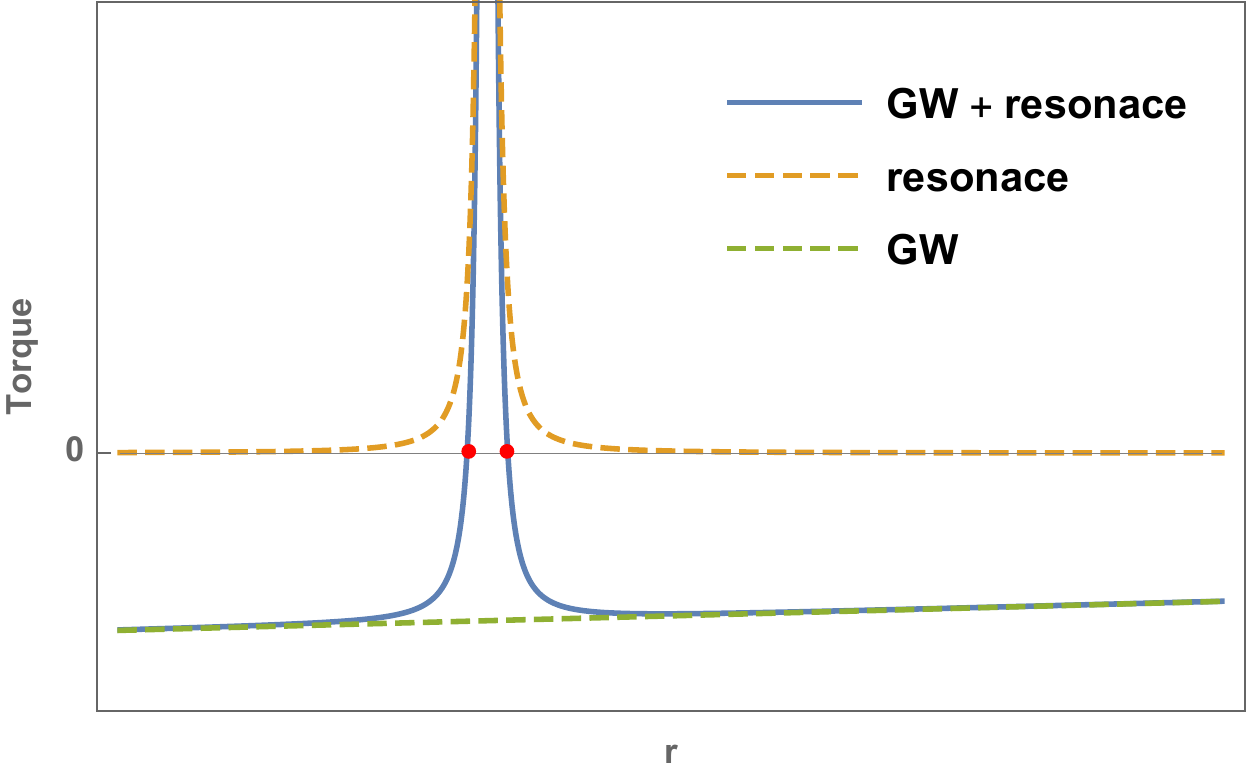}
\caption{The torques on the companion star caused by GW radiation (green dashed line), AM transfer (orange dashed line) and both (blue solid line). The two red dots show where the first two torques balance.}\label{fig:torque}
\end{figure}

\section{Astrophysical Implications}

According to Eq.~(\ref{eq:alpha}), if Axion(s) does exist around the $\mathcal{O}(10^{-17}-10^{-13})$eV range, it is possible to find astrophysical BHs with size comparable to the Axion Compton wavelength. For these systems, a co-rotating EMRI generically stalls  at the floating orbit instead of inspiralling into the central BH. It means that a supermassive or an intermediate mass BH (IMBH)  that matches the Axion mass, an EMRI may exist at all time until the could depletes, which may take Hubble time \footnote{The average capture time of an EMRI is $\sim \mathcal{O}(\rm Myrs)$, which is much shorter than the cloud lifetime. So we expect at least one EMRI floating around the central BH, if the Axion Compton wavelength matches the BH mass.}. As only co-rotating orbits are floated, for SMBHs within the right mass range, one may expect half of the EMRIs will be affected by floating orbit.
Note that although the observational evidences of IMBHs (see \cite{vanderMarel:2003ka,coleman2004intermediate}  for  reviews on intermediate mass BHs and the references therein for more detailed discussion) are still subject to debate, there are tentative implications by extrapolating the observed relation between the supermassive BH mass and its host galaxy mass \cite{gultekin2009fundamental,graham2011expanded,mcconnell2013revisiting,graham2013m}. Searching for IMBHs has been an active area of research so far, especially with the recent search using gravitational waves \cite{abbott2017search}.
For these IMBHs, the frequency of GWs from the floating orbits are possibly detectable by LISA. 
Fig.~\ref{fig:SNR} shows the signal-to-noise ratio \cite{Ruiter:2007xx} of GWs from floating orbits that may exist around sample observed intermediate mass BH candidates in the local group \cite{Pasham:2015tca, Maillard:2004wi, Feldmeier:2013gba, Ibata:2009nh, Anderson:2009ku}. 
Such observation will fill the mass gap  left  by observing GW radiation direct from the cloud \cite{Brito:2017wnc, Brito:2017zvb}.

\begin{figure}[tbp]	
\centering
\includegraphics[width=0.4\textwidth]{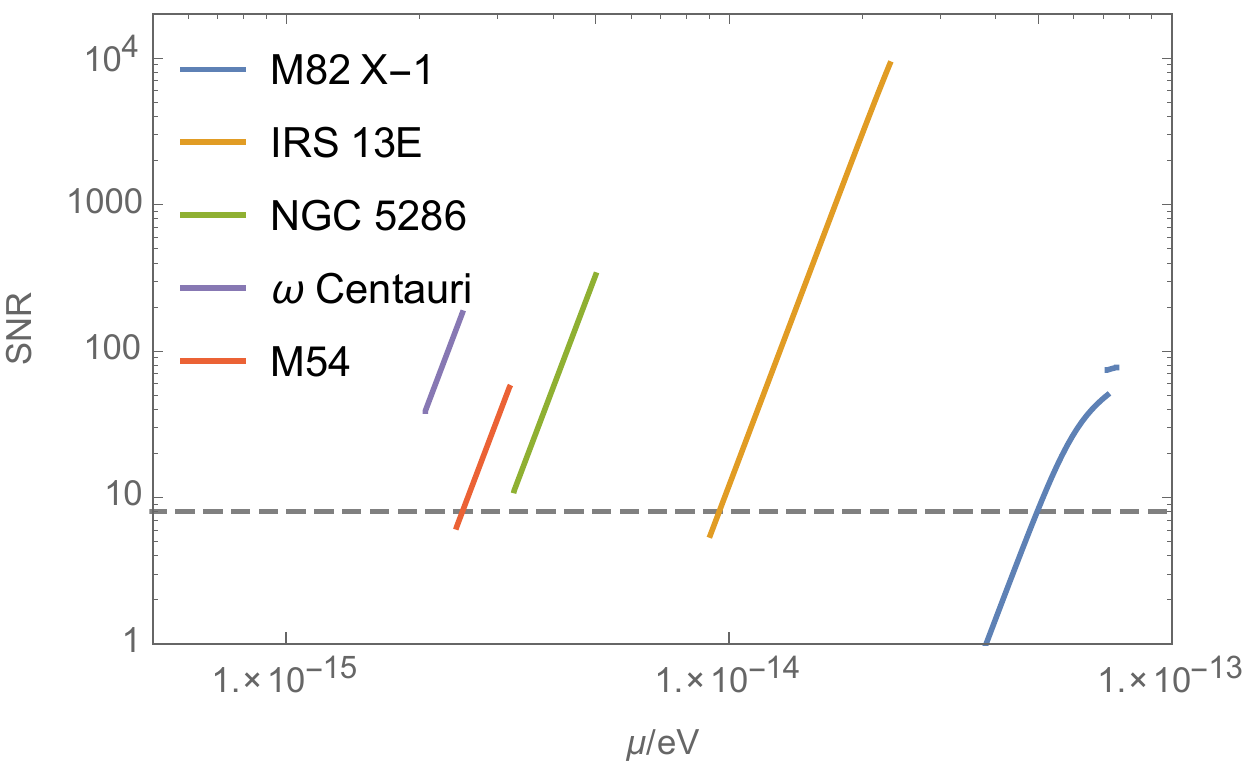}
\caption{Signal-to-noise ratio (SNR) of GW signals from floating orbits that could exist around some observed intermediate mass BH candidates in the local group, assuming a coherent observation time of 4 yrs for LISA. The mass (subject to measurement uncertainties) and distance ($M/M_\odot, D_L/\text{kpc}$) of the showing candidates (from top to bottom) are $\left(400, 3.6\times10^3\right)$\cite{Pasham:2015tca}, $\left(1300, 8\right)$\cite{Maillard:2004wi}, $\left(6000, 11\right)$\cite{Feldmeier:2013gba}, $\left(1.2\times10^4, 5\right)$\cite{Anderson:2009ku}, and $\left(9400, 26.3\right)$\cite{Ibata:2009nh} respectively.}\label{fig:SNR}
\end{figure}
%

With floating orbits, we may find much more in-plane EMRIs than expected, as the GW radiation will damp out the orbital AM on the equatorial plane of a inclined (and co-rotating) orbit, leaving the piece orthogonal to the plane supported by the cloud AM transfer. 
In addition, for a given supermassive BH, EMRI merger happens once per a few million years on average \cite{gair2017prospects}, depending on the mass of the supermassive BH. This could be much shorter than the lifetime of a floating orbit, such that by the time the second EMRI object enters the vicinity, the first EMRI object is still trapped at the floating orbit. Therefore 
it is possible to have multiple stellar-mass objects accumulating at comparable radius to the central BH, the mutual gravitational interaction between which may lead to  very interesting phenomena. 

For example, similar to planetary systems, these stellar-mass objects may experience Kozai-Lidov resonance \cite{kozai1962secular,lidov1962evolution,yang2017general}. They could also be locked into mean-motion resonances \cite{souchay2010dynamics}, with orbital frequencies being commensurate with each other. On the other hand, if the mean-motion resonance does not succeed, as the second EMRI object also has to across the 
floating resonance, and because of the migration to the equatorial plane, it is possible for it to scatter with the first EMRI object. This gravitational scattering may lead to 
the ejection of an EMRI object, and/or kick one EMRI object to a tighter orbit off the floating resonance. Moreover, it may result in a gravitational capture instead of scatter to form a stellar-mass BH binary, which undergoes the Kozai-Lidov resonance in the tidal field of the supermassive BH and mergers quickly. This stellar-mass binary merger produces gravitational waves in the LIGO band, and a heavier final BH most likely trapped in the floating orbit. The chance to have BH kicks to be comparable to the orbital speed of the centre of the mass of the binary, which is several percent of the speed of light, is rather insignificant \cite{gerosa2018black}. If this process is able to repeat many times during the lifetime of the supermassive BH, an intermediate mass BH may form from these mergers.  Theoretically assessing the likelihood and initial condition for different scenarios require long-term numerical integration for the orbital evolution of this multi-body system under gravitational interaction. The discussion of multi-body effects will be presented in a separated study.

%
%
%
\vspace{0.2cm}

{\it Acknowledgements-} We thank Sam Dolan and Horng Sheng Chia 
for inspiring discussions which initiated this work. We also thank Gongjie Li for discussion on the multi-body dynamics. J.Z. and H.Y.~acknowledge support from the Natural Sciences and Engineering Research Council of Canada, and in part by the Perimeter Institute for Theoretical Physics.  Research at Perimeter Institute is supported by the
Government of Canada through the Department of Innovation, Science and
Economic Development Canada, and by the Province of Ontario through
the Ministry of Research and Innovation. J.Z. is also supported by European Union's Horizon 2020 Research Council grant 724659 MassiveCosmo ERC-2016-COG.

\bibliography{master}
\end{document}